\documentclass[9pt,twocolumn,twoside]{opticajnl}
\journal{opticajournal} 

\setboolean{shortarticle}{false}

\usepackage{siunitx}
\usepackage{lineno}
\usepackage{stfloats}
\usepackage{comment}
\usepackage{array}
 \usepackage{setspace}
 \usepackage{xr}
\title{Packaged Cryogenic Photon Pair Source Using an Effective Packaging Methodology for Cryogenic Integrated Optics}

\author[1,2,4]{Donald Witt*}
\author[1,2]{Lukas Chrostowski}
\author[2,3]{Jeff Young}

\affil[1]{Department of Electrical and Computer Engineering, The University of British Columbia, V6T-1Z4, Vancouver, British Columbia, Canada}
\affil[2]{Stewart Blusson Quantum Matter Institute, V6T 1Z4, Vancouver, British Columbia, Canada}
\affil[3]{Department of Physics and Astronomy, The University of British Columbia, V6T 1Z1, Vancouver, British Columbia, Canada}
\affil[4]{John A. Paulson School of Engineering and Applied Sciences, Harvard University, Cambridge, MA 02138}

\affil[*]{donaldwitt@g.harvard.edu} 

\begin{abstract} 
A new cryogenic packaging methodology that is widely applicable to packaging any integrated photonics circuit for operation at both room temperature and cryogenic temperature is reported. The method requires only equipment and techniques available in any integrated optics lab and works on standard integrated photonic chips. Our methodology is then used to enable the measurement of a single photon pair sourced based on a silicon ring resonator at cryogenic temperatures. When operating at 5.9 K, this source is measured to have a peak pair generation rate 183 times greater then at room temperature in the CL-band. 
\end{abstract}
\setboolean{displaycopyright}{false} 
\begin{document}
\maketitle

\section{Introduction}
In the field of quantum information science, many integrative photonic devices need to operate at cryogenic temperatures. For example, detectors such as superconducting single wire nanowire detectors\cite{snspd1,snspd2} transition edge detectors \cite{tepd}, color emitters such as color centers such as \cite{wcenter,gcenter,tcenter,tcenteredo,tcenteredo2} and qunatum dots \cite{quantumdot} all operate in the cryogenic temperature regime. It is desirable to have a method of packaging such devices  at room temperature so that they can operate at both room temperature and cryogenic temperature without the need for any  realignment or adjustment. However, it is difficult to develop such packaging techniques due to the extreme temperature difference and the thermally induced stresses caused by the variation in thermal expansion between various materials. Due to stresses induced by this extreme temperature difference, many existing methods experience increased loss or fail completely when operating at cryogenic temperature \cite{gluegc1,cryogc2,cryogcfailed}. In addition, many existing techniques require specialized equipment for assembly \cite{cryopwb,cryogenictaper}. Several of these coupling methods are done one fiber at a time, limiting their scalability \cite{gluegc1,cryogc2,cryogenictaper}. A new method is required that can address the challenges of temperature related stresses, the need for specialized equipment, and scalability.

Here, we present a new methodology that packages photonic devices with low loss, and that simultaneously connects multiple fibers, while relying on equipment available in a standard optics lab. Using this new methodology, we are able to package chips with low loss across multiple fibres that can operate over temperatures ranging from room temperature to 5.9K. To demonstrate the applicability of our technique, we measured the single photon pair generation rate of a silicon photonic ring resonator pumped around 1560 nm while operating at both room temperature and 5.9K. We showed that this ring resonator at cryogenic temperature has a 183 times greater photon pair generation rate and an improved coincidence to accidental count ratio (CAR) due to a large reduction in undesirable non-linear thermo-optic mode shifting in contrast to the kerr non-linearity on which the source operates. Our method is highly scalable, as it is compatible with widely available 64 channel fiber arrays.

\section{Packaging Methodology}
Grating couplers are commonly used to couple light to an integrate photonic circuits. They do this through a diffraction grating etched into the surface of the chip that is designed to diffract out of plane light from a fiber into an in-plane waveguide mode \cite{normphcgc,phcgcap,subwave,fabintheloop}. Grating couplers allow for the rapid characterization of many photonic components, as they can be placed anywhere on the surface of a chip, in contrast to techniques such as edge couplers, taper fiber coupling and photonic wirebonding, \cite{cryoedgecoupler} which only allow light to be input on the edge of a chip. Moreover, grating couplers do not require specialized equipment to fabricate, in contrast to photonic wirebonding and taper fiber coupling \cite{cryopwb,cryogenictaper}. The primary disadvantage of grating couplers are their high insertion loss and low bandwidth. However, this disadvantage can be overcome through careful design of the grating couplers \cite{fabintheloop}.

At cryogenic temperatures, the primary disadvantage of grating couplers is their need for precise alignment. The motorized stages used for this precise alignment do not operate well at cryogenic temperatures and are extremely expensive as well as prone to failure; thus, they are not present in every cryostat. Moreover, cryogenic experiments are done to minimize the sources of noise. To achieve this goal, many cryostats are completely enclosed without the presence of viewports. This means that the initial alignment needs to be done at room temperature with the cryostat open so that the alignment can be obtained visually using a microscope. During the cool down process, thermal contraction causes significant drifts in alignment, potentially leading to extended time spent attempting to reacquire the alignment blindly.

The challenges presented by cryogenic operation can be overcome with a packaging methodology that fixes the fiber array to the chip with the correct alignment. Such a fixed packaging methodology will eliminate the need for alignment inside the cryostat and the associated motorized stages. This methodology allows for the entire assembly to be made at room temperature in normal atmospheric conditions where the required alignment is much easier to perform.
However, the extreme thermal stress presented by cooling down from 295 K to 4 K or below has resulted in many such packaging methods experiencing increased insertion loss at low temperature.

\begin{figure*}[th!]
\centering
     \def\svgwidth{1\textwidth}
     \setstretch{1}
     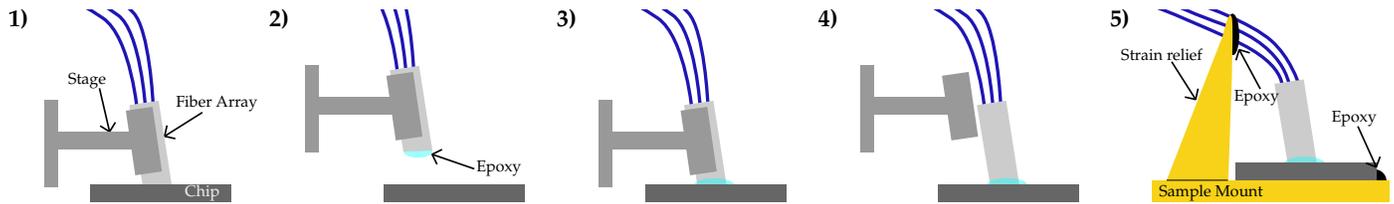
      \caption{The five steps required to package the device. 1) Find and measure the device using a a room temperature motorized stage and laser. 2) Raise the fiber array and apply the EP29LPSP epoxy. 3) Lower the array back into contact with the device. The epoxy is then allowed to cure for 24 hours followed by heat curing at 65C for 8 hours. 4) Disconnect the fiber array from the stage and unmount the chip. 5) The chip is then mounted to the sample mount using Apiezon N vacuum grease applied to the bottom of it. The chip is then secured using a drop of EP29LPSPAO-1 black epoxy on the edge. The fibers are secured to the strain relief using the same epoxy. This epoxy is then cured. The chip is now fully packaged. }
       \label{packaging_steps}
\end{figure*}

To overcome the extreme thermal stress, our method relies on three different techniques. First is the careful selection of the epoxy used during the assembly process. We select cryogenic compatible two-part epoxy EP29LPSP which is rated to work both at room temperature and cryogenic temperatures. While this epoxy takes longer to cure than UV cure epoxies, it does not experience cracking when exposed to cryogenic temperatures as has been commonly seen with UV epoxies \cite{uvfailure}. Secondly, we ensure that the epoxy is applied to the bottom of the fiber array. This ensures that the fiber array experiences minimal movement in relation to the grating coupler during cool down.
Third, we mount the sample to a copper sample mount that incorporates strain relief to avoid stressing the connection between the fiber array and the chip.
We follow the five steps as illustrated in Fig \ref{packaging_steps}.

\textit{Step one} is to characterize the device using the fiber array it will be packaged with.

\textit{Step two} is to apply the two-part epoxy to the bottom face of the fiber array. This is done by first raising the fiber array, mixing the two-part epoxy, and applying it to the face of the fiber array using the tip of a glass pipette. Care should be taken in this step to avoid scratching the face of the fiber array with the glass pipette. This can be seen in Fig. \ref{supp-glue_on_FA}. 

\textit{Step three} is to lower the fiber array into contact with the grating couplers. After the fiber array is lowered, the alignment is optimized by moving the stage. At this point in the process, the fiber array could be raised to move to a different device and the alignment reoptimized if desired. Once the alignment is optimized, a small drop of additional epoxy is added to the front of the array to ensure the strongest possible bond with the chip surface. The epoxy is then allowed to dry for 24 hours. The optical power passing through the device should be checked two hours into the drying process and the alignment fine-tuned if needed. During the drying process, the alignment can drift slightly, but it is possible to readjust it within this time frame. In our case this drift was found to be under 2 $\mu$m but is likely stage and temperature dependent. After drying at room temperature for 24 hours to gel the glue, the glue is heat cured at 65 C for 8 hours. This is done using the thermal electric cooler (TEC) present on the stage. If a TEC is unavailable it is also possible to transfer the sample to a hot plate for this cure.

\textit{Step four} is to remove the fiber array from the stage. In our setup, this is done by loosening two set screws. The fiber arm is then moved back using the motorized stage and the chip removed with the fiber array now bonded to the surface.

\begin{figure}[h!]
\centering
     \includegraphics[width=0.5\columnwidth]{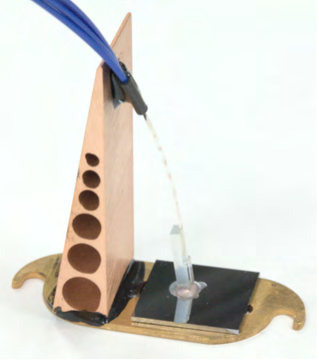}
      \caption{ A photograph of a packaged sample. Observe that the gold plated copper sample mount is at the base. Mounted to the sample mount, using a small drop of epoxy, is the silicon photonic chip. On the left is a copper block acting as strain relief for the fiber array. The fiber array is mounted to the silicon photonic chip using the methodology described in Fig. \ref{packaging_steps}.}
       \label{sample}
\end{figure}

\textit{Step five} is to attach the chip to the copper sample holder. This is done by applying Apiezon N vacuum grease to the bottom of the chip. Then the chip is placed on the copper sample holder and Kapton tape is used to secure it temporarily. Then a small drop of EP29LPSPAO-1 black epoxy is applied to one side. It is critical that this epoxy is only applied as a small drop on one side, as the difference in thermal contraction between the chip and copper will apply extreme stress to the chip if it is put on multiple sides. Kapton tape is used to secure the fiber temporarily to the strain relief while monitoring the optical power passing through the device. It is important to monitor the optical power during this process as strain on the fibers can increase the insertion loss. Epoxy is then used to secure the fibers to the strain relief. The epoxy is then allowed to dry at room temperature for 24 hours. It is then heat cured at 80C for 6 hours. This is done on a hot plate with a crystallization dish covering the packaged sample. The packaged sample is now ready for use. A photograph of a fully packaged sample can be seen in Fig. \ref{sample}.

\section{Insertion Loss}
\begin{figure}
\centering
     \def\svgwidth{1\columnwidth}
     \scriptsize
     \setstretch{1}
     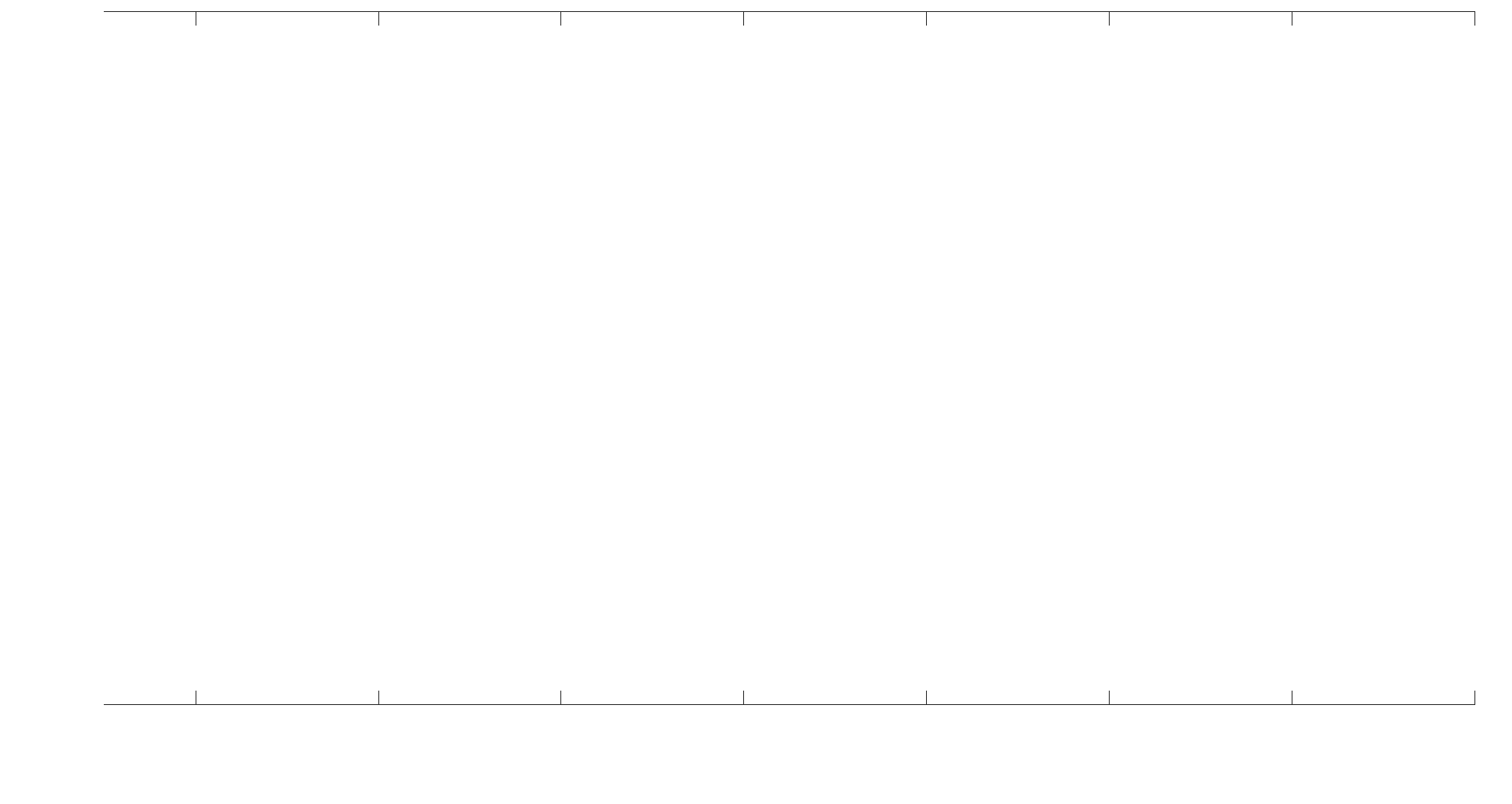
      \caption{A transmission measurement of the ring resonator insertion loss at both room temperature and 5.32 K. There was a negligible difference in the peak insertion loss between the two temperatures.}
       \label{packaged_gc_response}
\end{figure}

The key test of the effectiveness for any packaging technique is the insertion loss. This includes any increase in the insertion loss both from the packaging itself and from cooling down to cryogenic temperatures. To characterize the low insertion loss of our methodology, a silicon photonic chip was used. To minimize our starting insertion loss, a Fab-in-the-Loop optimized grating coupler was used \cite{fabintheloop} with an initial loss of  3.76 dB at 1618 nm. As our chip was air clad, meaning the epoxy would be in direct contact with the grating coupler, the starting grating coupler was biased by 14 nm to compensate for the shift caused by the epoxy. This shift can be seen in supplement Fig. \ref{supp-shiftglue}. The sample was then packaged using our packaging methodology. The sample was then placed in a custom-built closed cycle 4K cryostat. The insertion loss was then measured using a Keysight 81608A low SSE tunable laser and 81635A InGaAs photodetector at both room temperature and 5.3 K. The measurement spectrum can be seen in Fig. \ref{packaged_gc_response}. 

As expected, the peak coupling wavelength shifted as the device was cooled to cryogenic temperatures. However, the insertion loss only increased by approximately 0.43 dB. This negligible increase in loss and robustness to cryogenic operation was seen across multiple samples (see supplement Fig. \ref{supp-multisample}). These results are a significant improvement over previous cryogenic fiber to grating coupler packaging methodologies \cite{gluegc1,cryogc2}.   

\section{Single Photon Pair Source}
One useful application of such a packaged silicon photonic device is the production of single photon pairs. Such single photon pair sources can be produced by using a ring resonator to enhance the spontaneous four wave mixing (SFWM) in silicon due to its third order non-linearity \cite{fwm1,fwm2}. However, when pumped near 1560 nm at room temperatures, such sources have a low efficiency due largely to strong two photon absorption that both broadens and thermo-optically induced shifts the ring resonances \cite{roomtemparature_study}.

\begin{figure}[th!]
\centering
     \def\svgwidth{\columnwidth}
     \footnotesize
     \setstretch{0.65}
     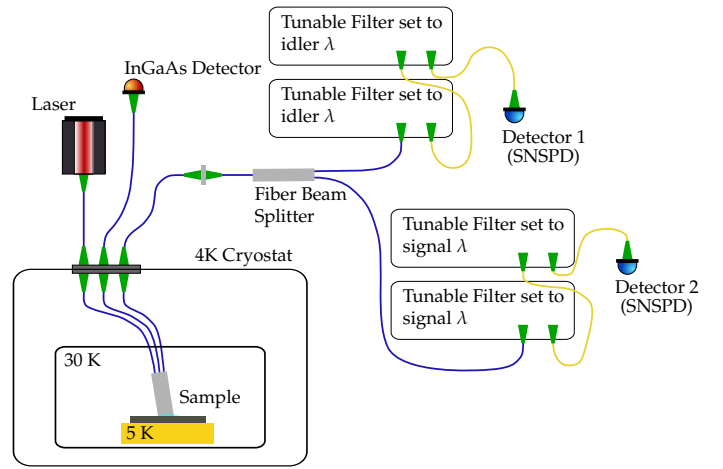
      \caption{An illustration of the optical path for the single photon measurements. A Keysight 81608A low SSE tunable laser was input into the silicon ring resonator. The signal from the through port of the ring resonator was then input into a Keysight 81635A InGaAs photodetector in order to monitor the transmission. The signal from the drop port of the ring resonator was input into a fiber beam splitter. The two outputs of the fiber beam splitter were then input into two sets of cascaded JDS Uniphase TB9 tunable optical filters. Each set of optical filters were tuned to the signal and idler wavelengths respectively. The filter outputs were then input into a set of Photon Spot superconducting nanowire single photon detectors (SNSPD). The electrical signals from these detectors were then input into a Swabian time tagger.}
       \label{setup_drawing}
\end{figure}

To measure the production of single photon pairs of such a source operating at cryogenic temperatures, one was packaged using our methodology. The source itself consisted of a 10 $\mu$m radius double bus ring resonator with a 240 nm coupling gap. This ring was packaged using our methodology by attaching a four channel polarization maintaining fiber array from Precision Micro Optics. The packaged source was then placed into our cryostat for both the room temperature and cryogenic measurements. The setup can be seen in Fig. \ref{setup_drawing}.

For the room temperature experiments, the signal filter was set to capture the photons generated at around 1554 nm. The idler filter bank was set to capture the photons generated at around 1572 nm. 

For the cryogenic experiments, the signal filter was set to capture the photons generated at around 1552 nm, and the idler filter was set to capture the photons generated at around 1571 nm.

\begin{figure*}
\centering
     \def\svgwidth{\textwidth}
     \scriptsize
     \setstretch{0.65}
     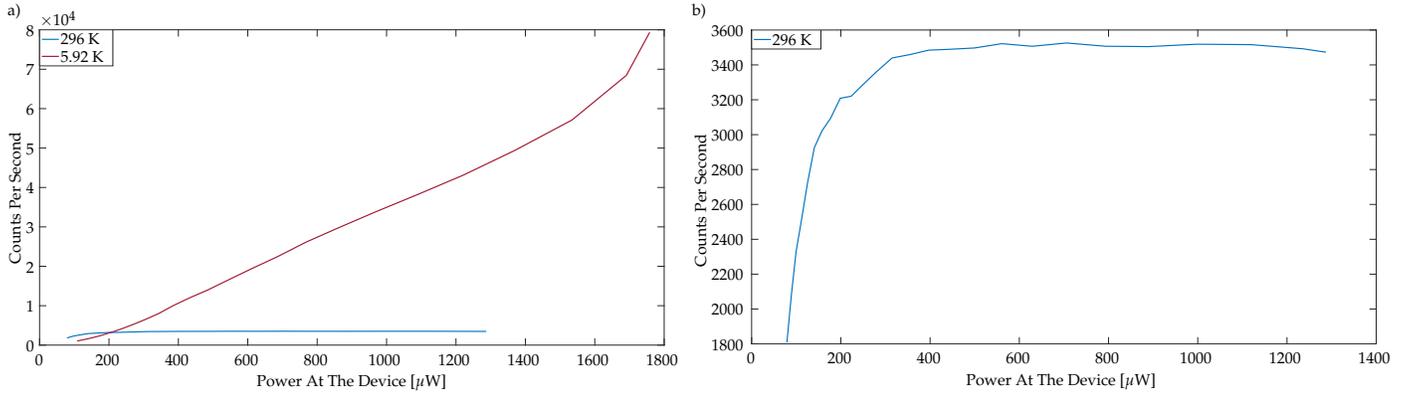
      \caption{a) A plot of the photon count rate versus power input into the device. The difference in loss between the two temperatures was accounted for in the pump power and by correcting the photon count rate in the room temperature case to account for the increased loss (see supplement section \ref{supp-losscal}). In red is the single photon count rate at 5.9 K. In blue is the single photon count rate at room temperature. When operating at cryogenic temperatures, the photon count rate saturated at much higher laser powers than at room temperature. b) A magnified plot of the room temperature count rate versus input laser power also corrected for the increased loss compared to cryogenic operation. One can see the count rate plateaus at approximately 318 $\mu$W.}
       \label{saturation_power}
\end{figure*}

Using this setup, the source was then characterized at both room temperature and 5.9k. First, the photon count rate on detector 1 was measured versus the input laser power. This was done to determine the power at which the photon count rate would saturate and thus determine the optimal operating point for both conditions. The pump laser was set to a wavelength of 1563.268 nm for the room temperature measurements and 1561.944 nm for the cryogenic measurements. These wavelengths correspond to the ring resonance between the signal and idler resonances. These results can be seen in Fig \ref{saturation_power}.

From these results, it is clear that the photon count rate at cryogenic temperature saturates at a much higher power than at room temperature. At room temperature, the photon count rate saturated at approximately 318 $\mu$W of power at the device. At 5.9K, the source has still not saturated at 1758 $\mu$W which is the maximum output power of our laser for this wavelength. 

The dramatic difference is qualitativley consistent with the 45\% reduction in the two photon absorption coefficient  \cite{cryotwophoton} versus only a 25\% reduction in the Kerr coefficient  \cite{cryotwophoton}, and the orders of magnitude reduction of the thermo-optic coefficient, from $10^{-4}$ $K^{-1}$ to $10^{-8}$ $K^{-1}$ \cite{thermooptic}, on going from room temperature to 5.9K. These differences in the non-linear effects are more directly manifest in the much greater threshold for the onset of peak dragging during swept laser scans (see supplement section \ref{supp-sweept_nonlinear}).

\begin{figure}[th!]
\centering
     \def\svgwidth{\columnwidth}
     \scriptsize
     \setstretch{0.65}
     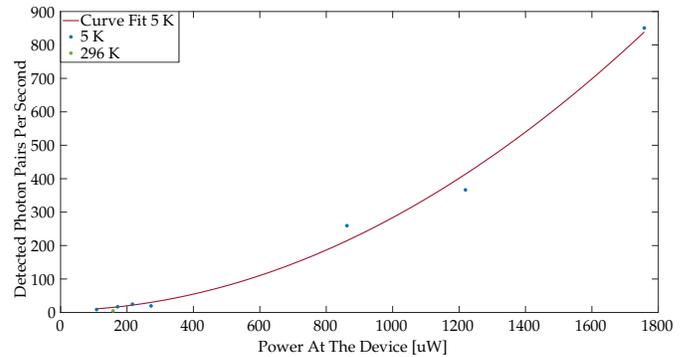
      \caption{ The photon pairs per second at cryogenic temperature in red with the data points in blue.  In green are the photon pairs per second at room temperature at 158 $\mu$W calibrated for the difference in loss between the two operating conditions (see supplement section \ref{supp-losscal}).}
       \label{pairgenrate}
\end{figure}

Based on these results, we measured the photon pair generation rate. To do this, correlation histograms were collected using a 10 picosecond bin size. Then a gaussian function was fitted to these histograms and the gaussian integrated over its full width half max (FWHM). This integrated value was divided by the collection time. These histograms were gathered over a 10 minute collection time while the source was operating at cryogenic temperatures. Due to the much lower pair generation rate at room temperature, the room temperature histogram was collected over a one hour collection time for comparison. The pair generation rate can be seen in Fig. \ref{pairgenrate}.

As expected from the results presented in Fig. \ref{saturation_power}, the photon pair generation rate is much greater when the source is operating at cryogenic temperatures. It also follows the general trends presented Fig. \ref{saturation_power} while operating at  both cryogenic and room temperature (see supplement  Fig. \ref{supp-pair_rate_room} for the room temperature pair generation rate plot). The pair generation rate versus power is represented by the fit given in Eqn. \ref{eqn-pairfit}. 

\begin{equation}
\frac{Pairs}{Second}=(2.587\times 10^{-4})P^2 + (1.912\times 10^{-2})P + 5.739
 \label{eqn-pairfit}
\end{equation}

Where P is the power at the device in $\mu$W. At cryogenic temperatures, the maximum measured photon pair generation of 850 pairs per second which is 183 times greater than the maximum room temperature rate after accounting for the difference in loss between the two operating conditions (see supplement \ref{supp-histogramlinear}). This can be explained but the lack of the saturation effect seen at low put powers while operating at room temperature. 

\begin{figure}
\centering
     \def\svgwidth{\columnwidth}
     \scriptsize
     \setstretch{0.65}
     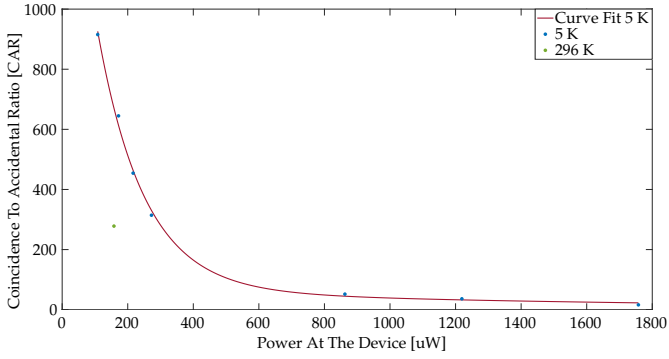
      \caption{The  coincidence to accidental count ratio (CAR) at cryogenic temperature in red with the data points in blue. In green is the room temperature CAR at 158 $\mu$W collected with a 1 hour collection time for comparison. }
       \label{carplot}
\end{figure}

The coincidence to accidental count ratio (CAR) was determined by fitting and integrating the peak of the correlation histogram in the same manner. Due to the low number of accidental counts the accidental count rate was determined as follows. First the average accidental count rate (AAC) was determined in areas away from any correlation peaks. Then this average was multiplied by the used the integrate the peak. Then Eqn. \ref{eqn-car} was used to calculated the CAR. 
\begin{equation}
CAR=\frac{\int_{{FWHM}} Peak - AAC \times  FWHM}{  AAC \times FWHM}
 \label{eqn-car}
\end{equation}
The CAR is also significantly improved when operating at cryogenic temperatures as seen in Fig \ref{carplot}.   

The CAR versus power is represented by the fit given in Eqn. \ref{eqn-carfit}. 

\begin{equation}
CAR=1846e^{-P\times 7.049 \times 10^{-3}}+ 71.23e^{-P\times 6.489 \times 10^{-4}}
 \label{eqn-carfit}
\end{equation}

Where P is the power at the device in $\mu$W. The CAR at a power of 158 $\mu$W of is 278 at room temperature. The CAR at a power of 158 $\mu$W while operating at cryogenic temperature is 670 or 2.41 greater then room temperature. The photon pair generation rate is at this power at cryogenic temperature is 15.22 pairs per second versus 4.63 pairs per second at room temperature or increase of 3.28 times.

\section{Conclusion}
Our new packaging method produces samples that operate at cryogenic temperatures. Although we demonstrated its utility for producing single photon pairs from a silicon ring resonator at cryogenic temperatures, our method can be used to package any quantum optical device. In addition to the large increase in performance secured by the capacity to operate at cryogenic temperatures, these sources can be integrated easily with other cryogenic components, such as single-photon detectors to form larger quantum systems. These results also show the potential gains in operating other non-linear silicon photonic devices at cryogenic temperatures. By further improving the grating coupler design the loss can be even further reduced (see supplement  Fig. \ref{supp-bestgc} for such an improved design). We anticipate that our method can also be applied to many other integrated photonic platforms such lithium niobate \cite{cryogenicln}, silicon nitride \cite{cryosin} and silicon carbide \cite{siccolor}.

\section{Acknowledgments}
This work is supported by the Natural Sciences and Engineering Research Council of Canada (NSERC), the B.C. Knowledge Development Fund (BCKDF), the Canada Foundation for Innovation (CFI) and the SiEPICfab consortium. We would like to thank Becky Lin for her help with the cryostat. We would like to thank Suzanne Smith for the writing advice. We would also like to thank the staff of the Stewart Blusson Quantum Matter Institute’s Advanced Nanofabrication Facility for their assistance.

\bibliography{cryopackaging}
\end{document}


\maketitle
\section{Glue Application}
\begin{figure}[H]
\centering
     \def\svgwidth{0.7\columnwidth}
     \setstretch{0.65}
\begingroup%
  \makeatletter%
  \providecommand\color[2][]{%
    \errmessage{(Inkscape) Color is used for the text in Inkscape, but the package 'color.sty' is not loaded}%
    \renewcommand\color[2][]{}%
  }%
  \providecommand\transparent[1]{%
    \errmessage{(Inkscape) Transparency is used (non-zero) for the text in Inkscape, but the package 'transparent.sty' is not loaded}%
    \renewcommand\transparent[1]{}%
  }%
  \providecommand\rotatebox[2]{#2}%
  \newcommand*\fsize{\dimexpr\f@size pt\relax}%
  \newcommand*\lineheight[1]{\fontsize{\fsize}{#1\fsize}\selectfont}%
  \ifx\svgwidth\undefined%
    \setlength{\unitlength}{610.93417911bp}%
    \ifx\svgscale\undefined%
      \relax%
    \else%
      \setlength{\unitlength}{\unitlength * \real{\svgscale}}%
    \fi%
  \else%
    \setlength{\unitlength}{\svgwidth}%
  \fi%
  \global\let\svgwidth\undefined%
  \global\let\svgscale\undefined%
  \makeatother%
  \begin{picture}(1,1.20975383)%
    \lineheight{1}%
    \setlength\tabcolsep{0pt}%
    \put(0,0){\includegraphics[width=\unitlength,page=1]{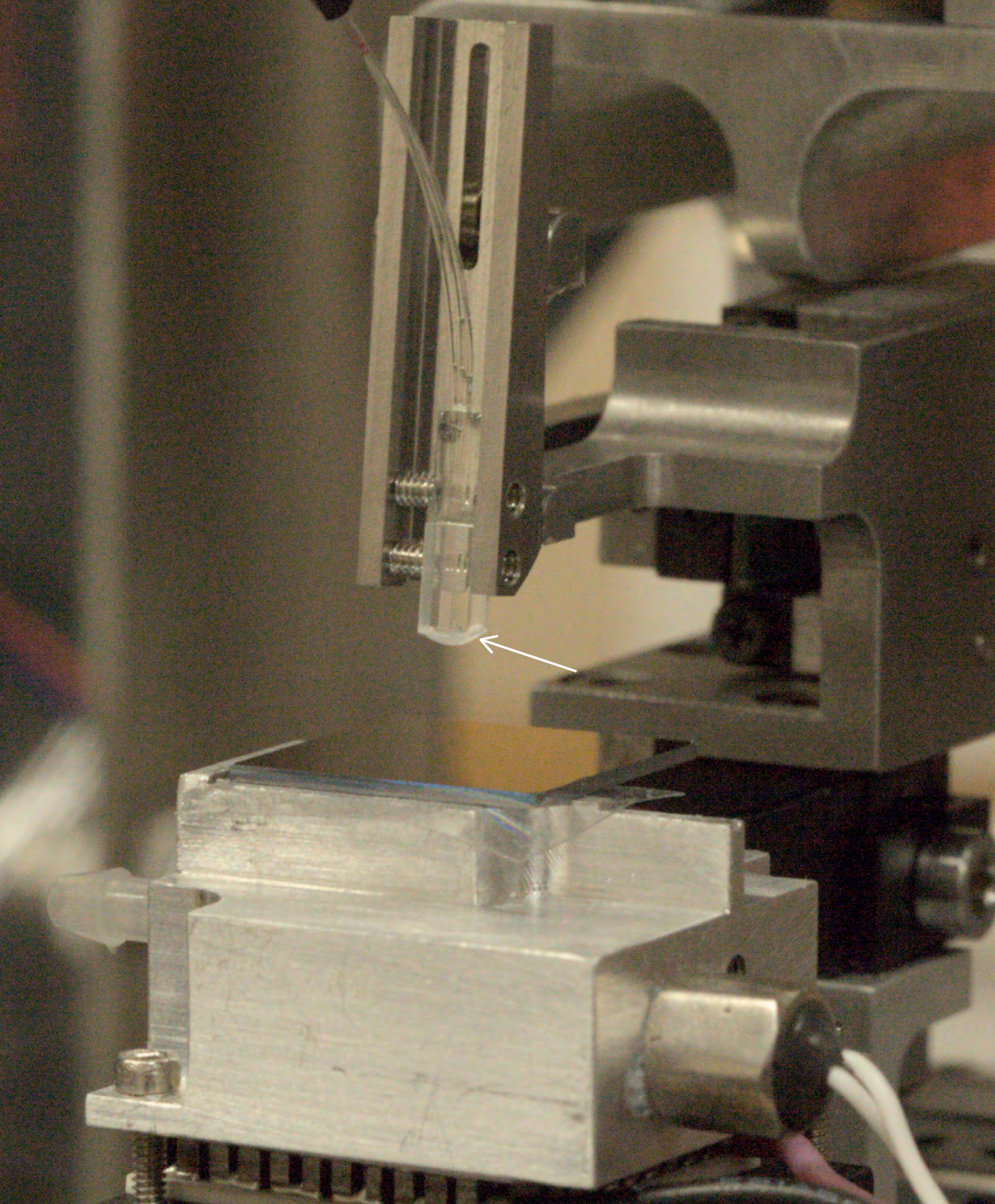}}%
    \put(0.62465525,0.52067878){\color[rgb]{1,1,1}\makebox(0,0)[t]{\lineheight{1.25}\smash{\begin{tabular}[t]{c}Glue\end{tabular}}}}%
    \put(0,0){\includegraphics[width=\unitlength,page=2]{glue_application.pdf}}%
    \put(0.15754715,0.49290478){\color[rgb]{1,1,1}\makebox(0,0)[t]{\lineheight{1.25}\smash{\begin{tabular}[t]{c}Chip\end{tabular}}}}%
    \put(0,0){\includegraphics[width=\unitlength,page=3]{glue_application.pdf}}%
    \put(0.233452,0.64253159){\color[rgb]{1,1,1}\makebox(0,0)[t]{\lineheight{1.25}\smash{\begin{tabular}[t]{c}Fiber Array\end{tabular}}}}%
  \end{picture}%
\endgroup%

      \caption{The drop of EP29LPSP epoxy can be seen on the fiber array before the array is brought into contact with the chip.}
       \label{glue_on_FA}
\end{figure}

\section{Shift Due To Epoxy}
\begin{figure}[H]
\centering
     \def\svgwidth{\columnwidth}
     \setstretch{0.65}
     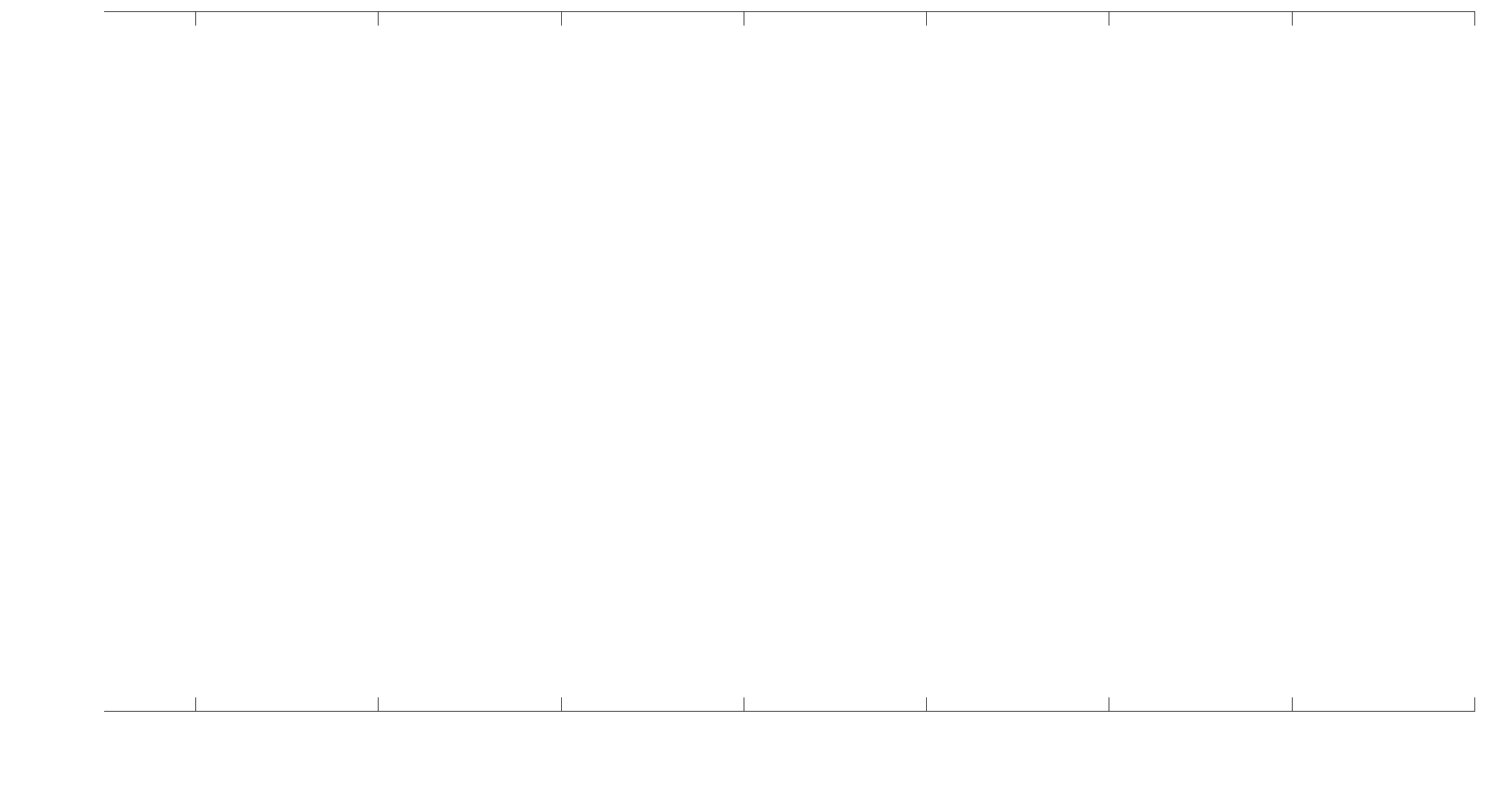
      \caption{The shift in the insertion loss spectrum caused by the application of the EP29LPSP epoxy. The peak wavelength shifted from 1552nm without epoxy to 1610nm with epoxy.}
       \label{shiftglue}
\end{figure}

\section{Multiple Sample Insertion Loss}
\begin{figure}[H]
\centering
     \def\svgwidth{\columnwidth}
     \setstretch{0.65}
     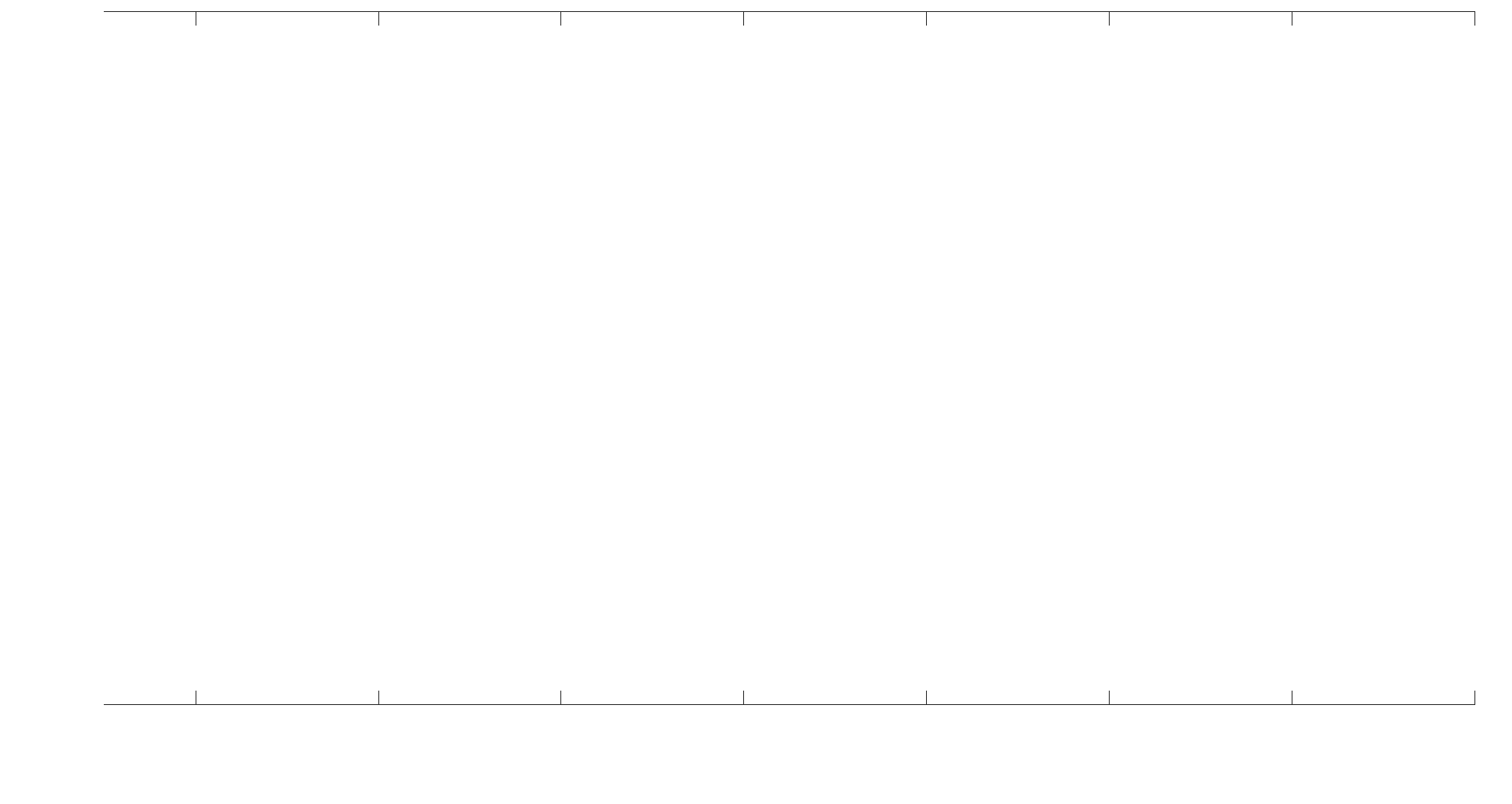
      \caption{The overlaid insertion loss spectrums of multiple packaged samples. Samples 1 and 2 were simple grating coupler loopbacks without a silicon micro-ring resonator. They utilized a more primitive version of the strain relief assembly that can be seen in Fig. \ref{sampleold}. This caused the strain relief to crack, damaging sample 2 while loading it in the cryostat. Despite the damage, it was still functional down to cryogenic temperatures. The improved strain relief assembly seen in the main text Fig. \ref{main-sample} eliminated this possibility.}
       \label{multisample}
\end{figure}

\section{Original Strain Relief Mechanism}
\begin{figure}[H]
\centering
     \includegraphics[width=0.5\textwidth]{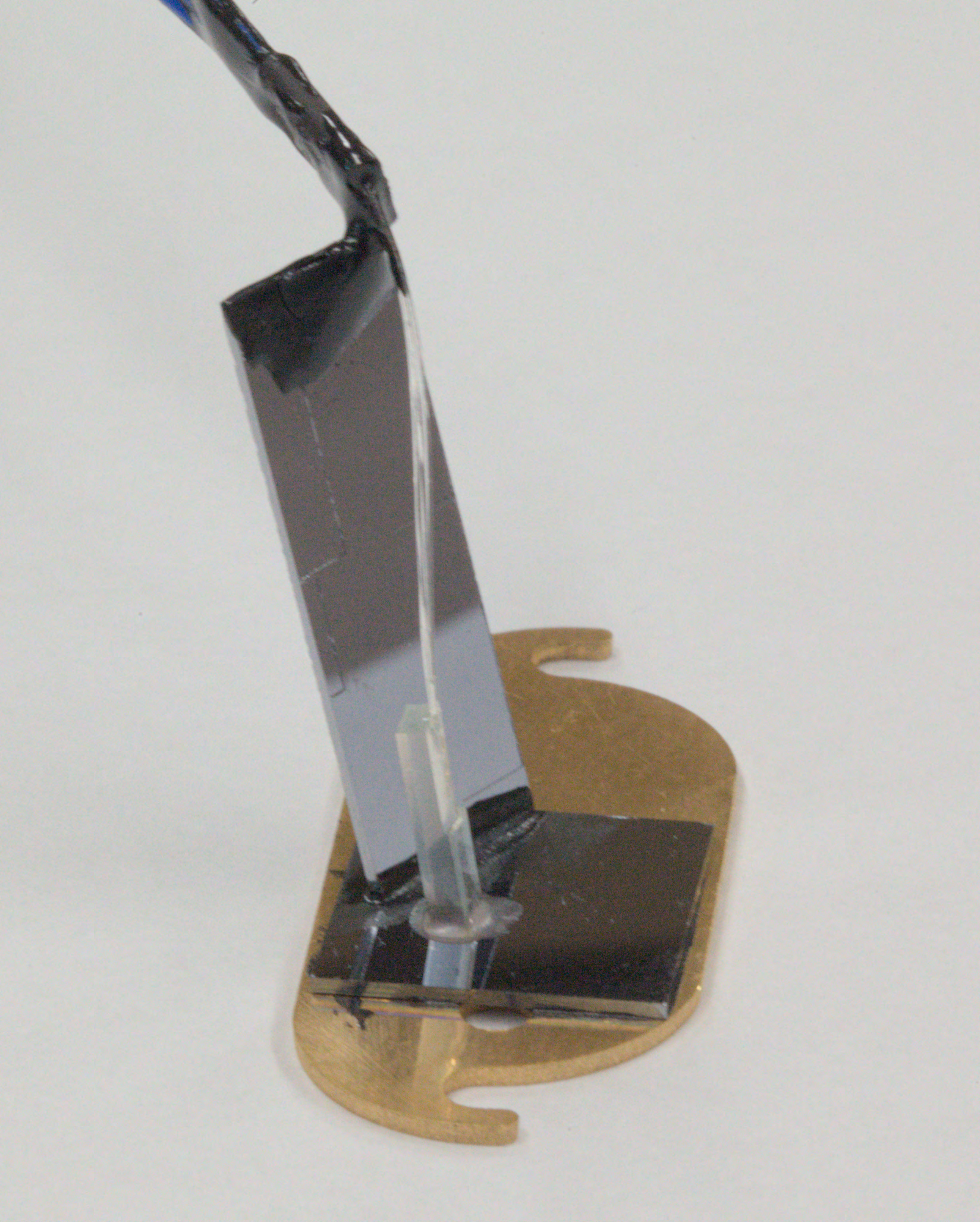}
      \caption{The original strain relief mechanism. This strain relief mechanize proved much less durable then our final strain relief mechanism, seen in main text Fig. \ref{main-sample} as it transferred strain directly to the chip potentially damaging it.}
       \label{sampleold}
\end{figure}

\section{Single Photon Loss Calculation}
\label{losscal}
At room temperature the pump loss was determined to be 11.33 dB by measuring the transmission and dividing by 2. Cryogenic pump loss was determined to be 9.96 dB by the same method.

The room temperature 1572 nm device and feedthrough photon loss was determined to be 9.31 dB. The filter loss was determined to be 11.02 dB measuring its transmission separately. The total loss at this wavelength was then calculated by adding 3 dB for the beam splitter plus an extra 3 dB for the photons sent to the wrong port for a total of 26.32 dB.

The room temperature 1554 nm device and feedthrough photon loss was determined to be 13.58 dB. The filter loss was determined to be 7.40 dB measuring its transmission separately. The total loss at this wavelength was then calculated by adding 3 dB for the beam splitter plus an extra 3 dB for the photons sent to the wrong port for a total of 26.98 dB.

The cryogenic 1571 nm device and feedthrough photon loss was determined to be 8.34 dB. The filter loss was determined to be 10.38 dB measuring its transmission separately. The total loss at this wavelength was then calculated by adding 3 dB for the beam splitter plus an extra 3 dB for the photons sent to the wrong port for a total of 24.72 dB.

The cryogenic 1552 nm device and feedthrough photon loss was determined to be 12.02 dB. The filter loss was determined to be 7.19 dB measuring its transmission separately. The total loss at this wavelength was then calculated by adding 3 dB for the beam splitter plus an extra 3 dB for the photons sent to the wrong port for a total of 25.21 dB.

In the 1570 nm range the room temperature condition has 1.605 dB of excess loss compared to the cryogenic case. In the 1554 nm range the room temperature condition has 1.773 dB of excess loss compared to the cryogenic case. For the histogram, the higher of these two differences was applied as a 1.504 multiplier on the room temperature data.

For the photon count comparison, the 1570 range multiplier of 1.447 was applied to the room temperature data. 

\section{Photon Pair Histogram}
\begin{figure}[H]
\centering
     \def\svgwidth{\columnwidth}
     \setstretch{0.65}
     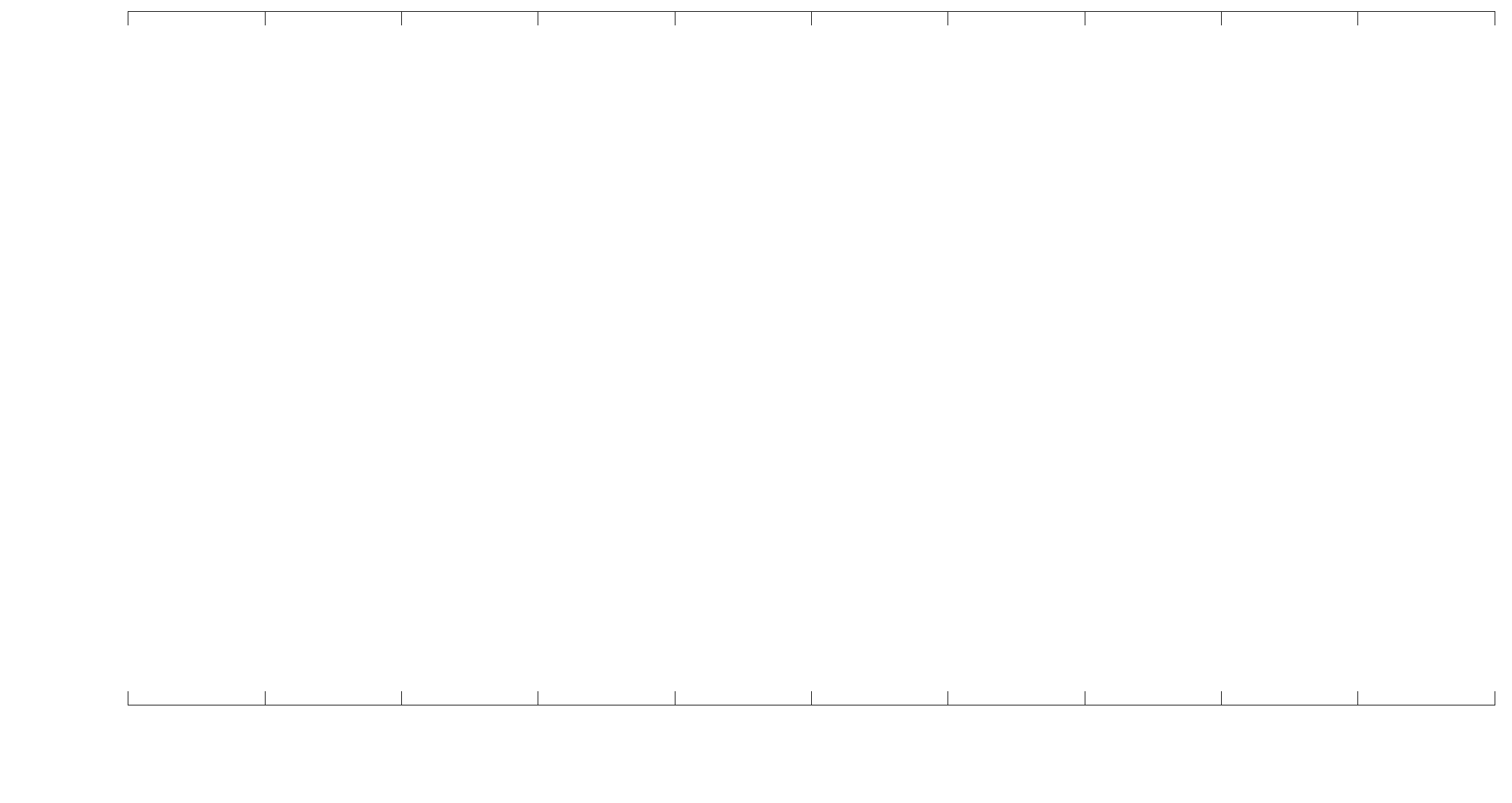
      \caption{A histogram of the single photon pairs generated by the ring resonator photon pair source. The difference in loss between the two temperatures was accounted for in the pump power and by correcting the photon count rate in the room temperature case to account for the increased loss (see supplement section \ref{losscal}). In red are the single photon pairs generated while operating at 5.9 K with a pump power of 1758 $\mu$W. In blue are the single photon pairs generated at room temperature with a pump power of 158 $\mu$W. The pair generation rate showed a large increase while operating at 5.9K compared to room temperature.}
       \label{histogramlinear}
\end{figure}

\section{Room Temperature Pair Generation Rate}
\begin{figure}[H]
\centering
     \def\svgwidth{\columnwidth}
     \setstretch{0.65}
     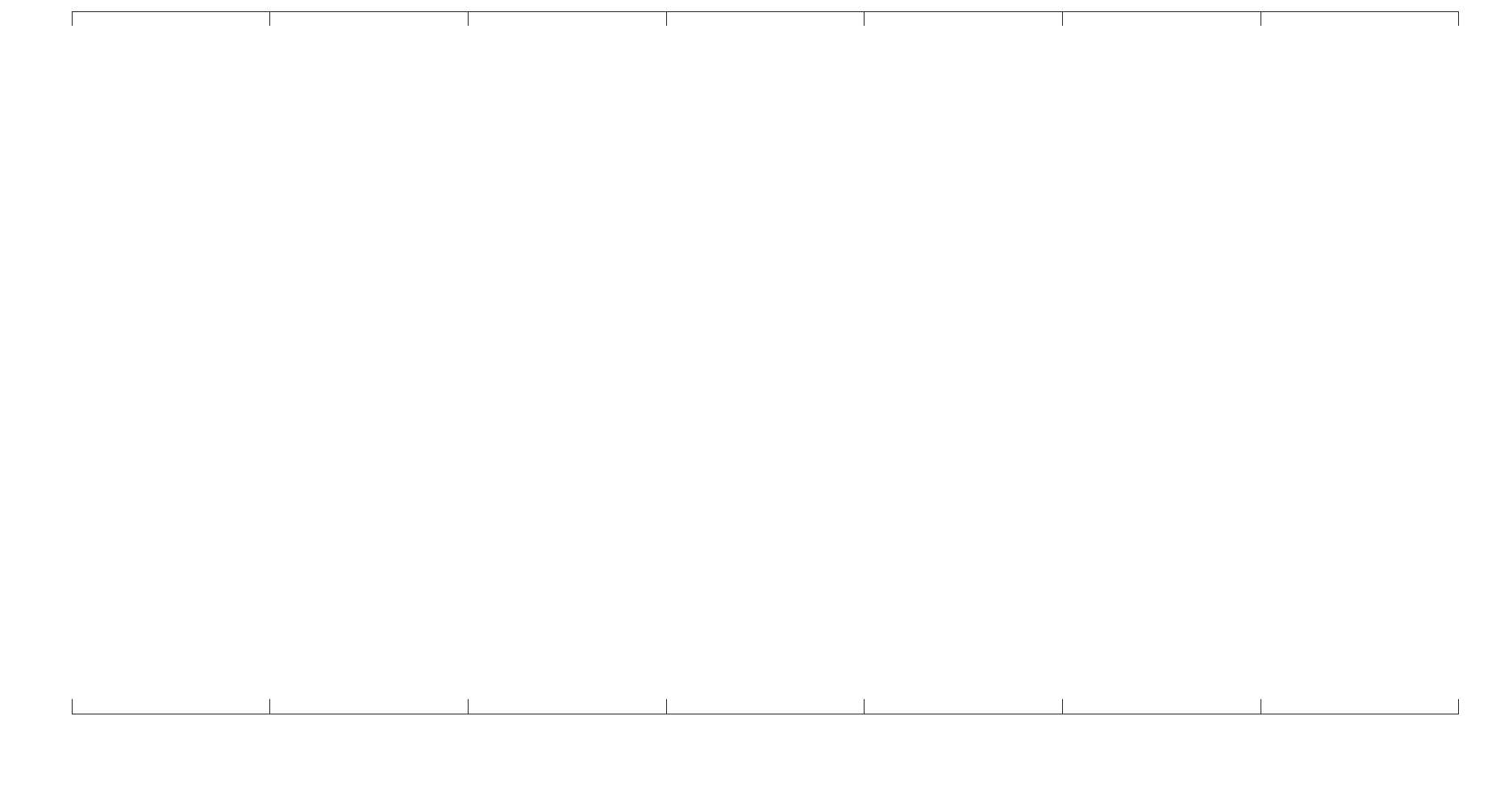
      \caption{The single photon pairs generated by the ring resonator photon pair source at room temperature versus pump power. The difference in loss between room temperature and cryogenic temperature was accounted for in the pump power and by correcting the photon count rate in this room temperature case to account for the increased loss (see supplement section \ref{losscal}). The trend follows the photon count rate plot seen in Fig. \ref{main-saturation_power}\color{blue}{b}. }
       \label{pair_rate_room}
\end{figure}

\section{Non-linear Optical Response}
\label{sweept_nonlinear}
\begin{figure}[H]
\centering
     \def\svgwidth{\columnwidth}
     \setstretch{0.65}
     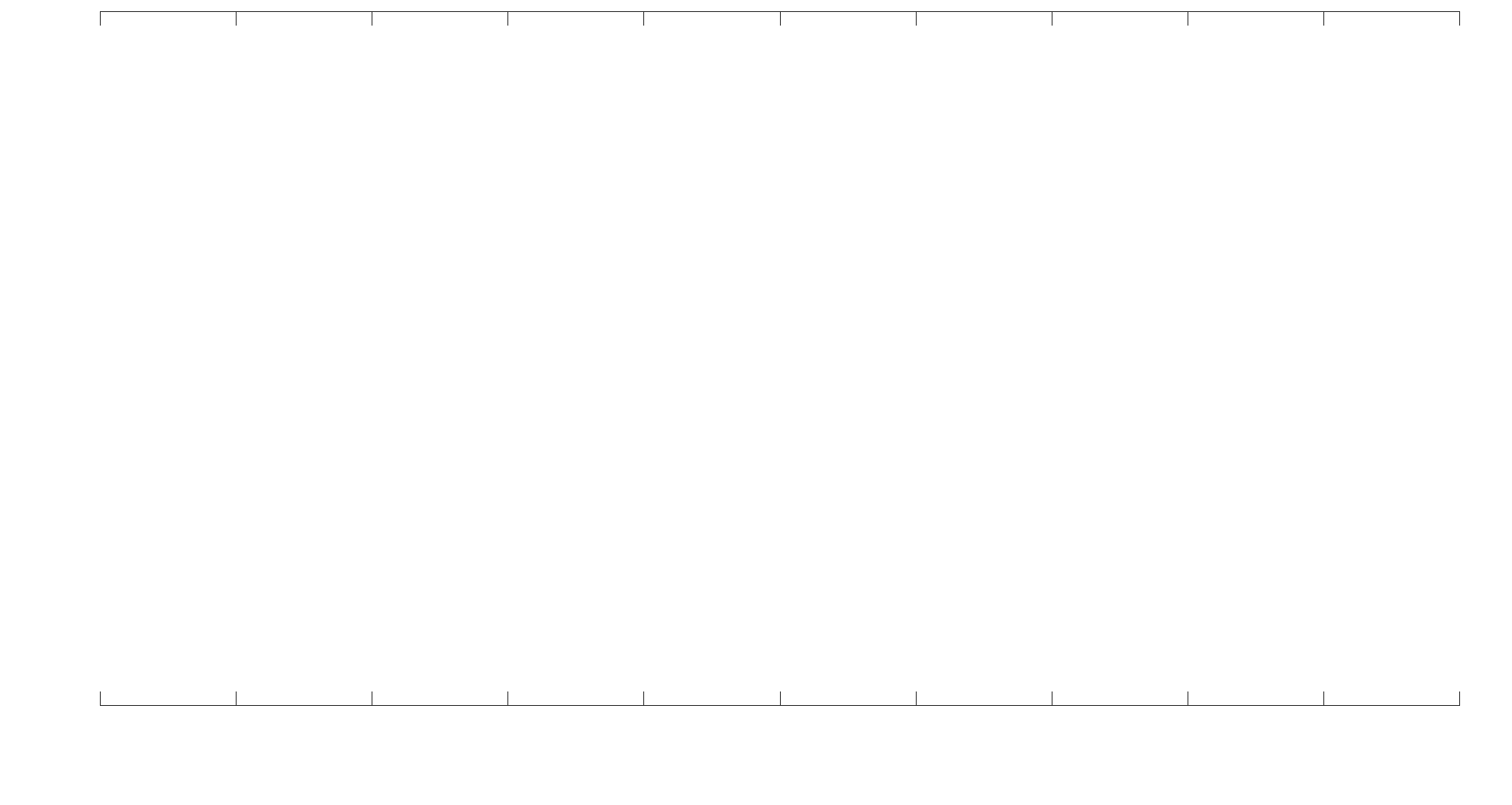
      \caption{A room temperature power series. One can see the visible non-linearity commencing at 59.05 $\mu$W of power at the device.}
       \label{room_nonlinear}
\end{figure}

\begin{figure}[H]
\centering
     \def\svgwidth{\columnwidth}
     \setstretch{0.65}
     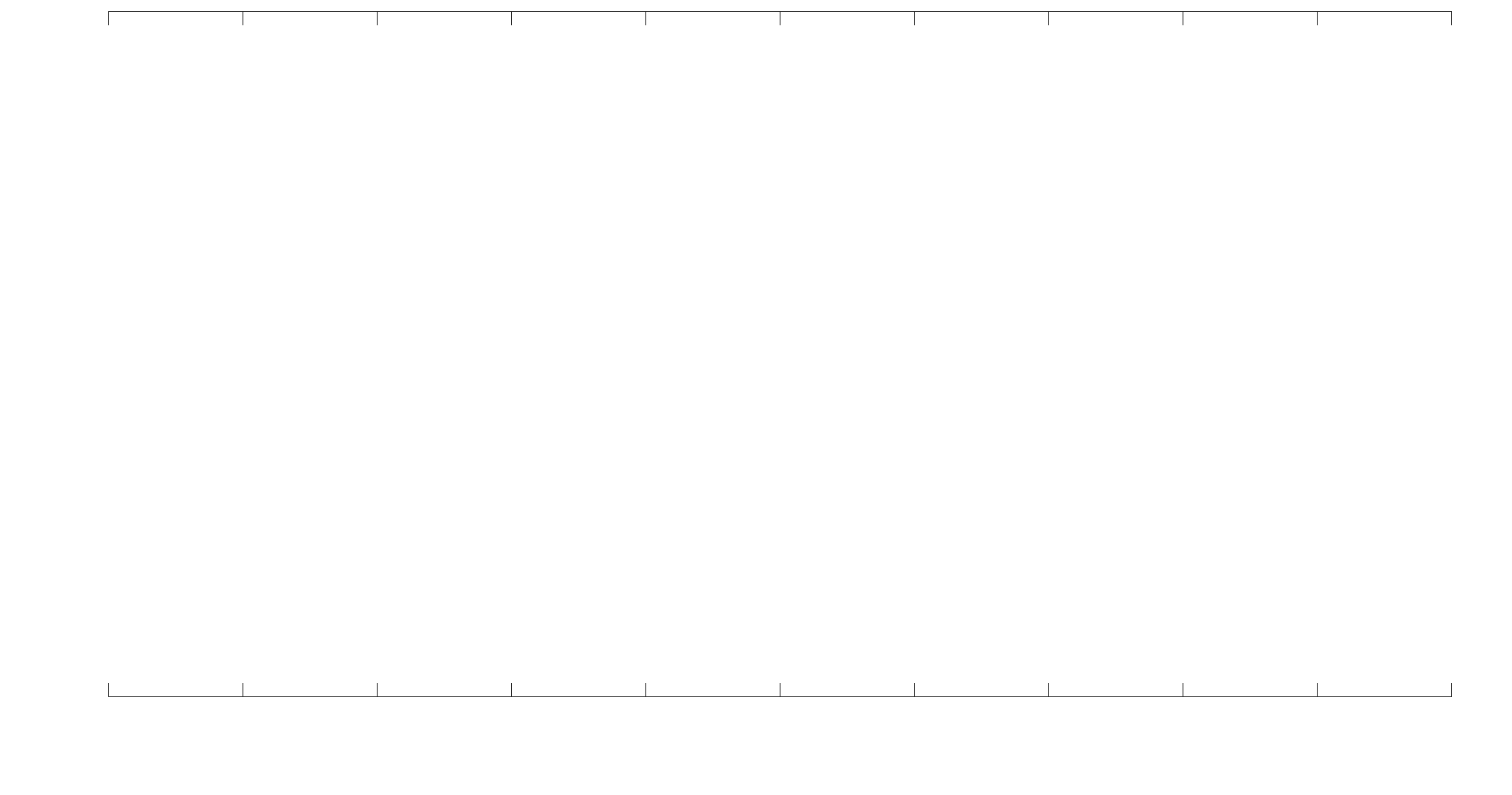
      \caption{A cryogenic power series at 5.9 K. One can see the visible non-linearity commencing at 2173.18 $\mu$W of power at the device which is much greater than the room temperature case seen in Fig. \ref{room_nonlinear}.}
       \label{cryo_nonlinear}
\end{figure}

\section{Improved GC}
\begin{figure}[H]
\centering
     \def\svgwidth{\columnwidth}
     \setstretch{0.65}
     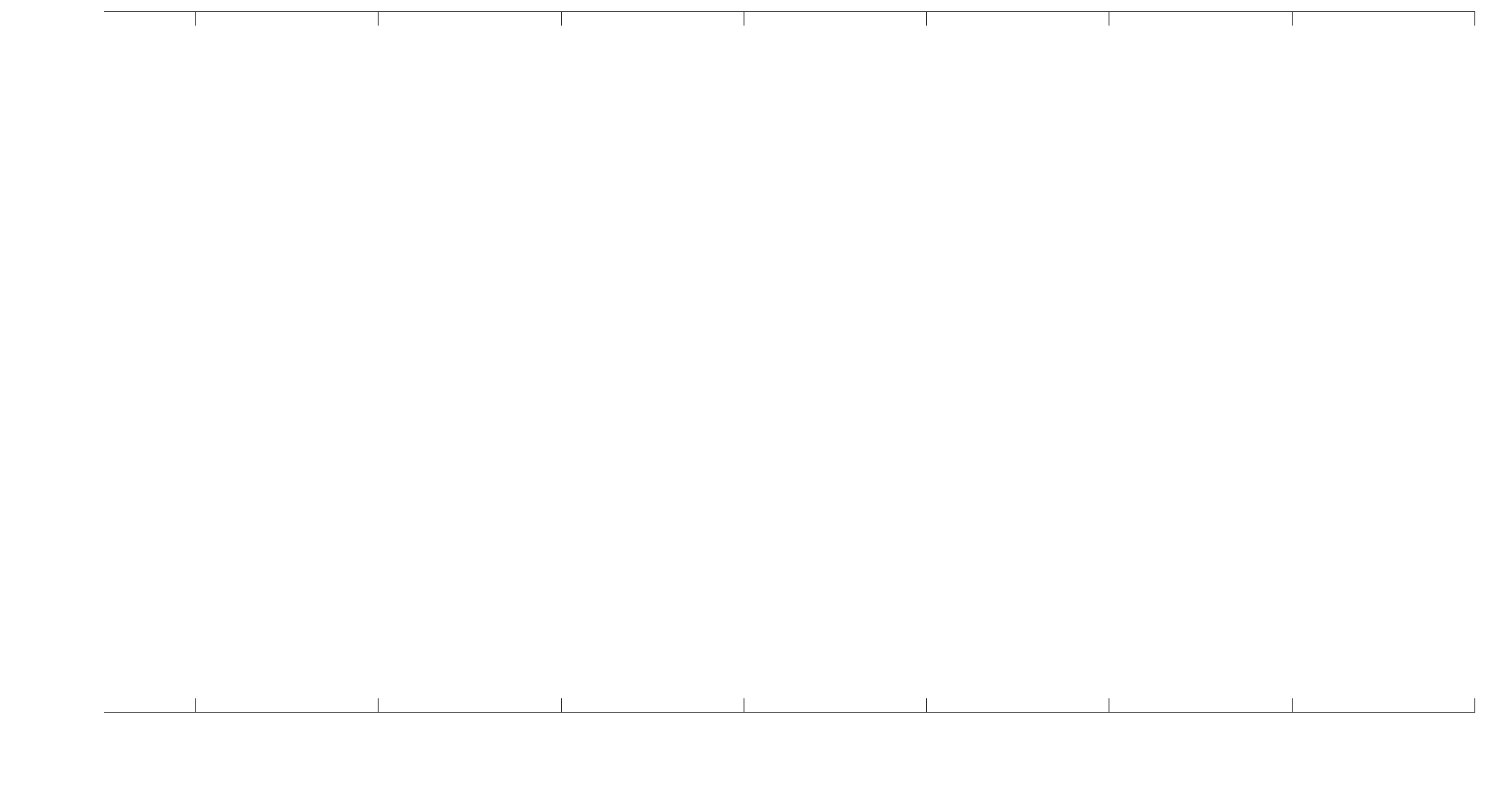
      \caption{A unpackaged single etch fab-in-the-loop optimized \cite{fabintheloop} grating coupler with only 2.40 dB of loss per coupler. By optimizing this coupler to work in the presence of the shift caused by the epoxy as seen in Fig. \ref{shiftglue}, it should be possible to package cryogenic devices with loss similar to cryogenic photonic wirebonding \cite{cryopwb}.}
       \label{bestgc}
\end{figure}

\bibliography{supplement}